\documentclass[]{spie}  %>>> use for US letter paper
\usepackage{amsmath,amsfonts,amssymb}
\usepackage{graphicx}
\usepackage[colorlinks=true, allcolors=blue]{hyperref}

\usepackage{mathtools}
\DeclareMathOperator*{\argmax}{arg\,max}

\title{A comparative study of methods to estimate conversion gain in sub-electron and multi-electron read noise regimes}

\author[a]{Aaron Hendrickson}
\author[b]{David P. Haefner}
\affil[a]{U.S. Navy, NAWCAD DAiTA Group, Atlantic Ranges \& Targets, Electro-Optical Tracking Systems, 23013 Cedar Point Road, Bldg. 2118, Patuxent River MD, 20670}
\affil[b]{U.S. Army Combat Capabilities Development Command (DEVCOM), C5ISR Center, Research \& Technology Integration (RTI)}

\authorinfo{Further author information: (Send correspondence to A.H.)\\A.H.: E-mail: aaron.j.hendrickson2.civ@us.navy.mil\\  D.P.H.: E-mail: david.p.haefner.civ@army.mil}

\begin{document} 
\maketitle

\begin{abstract}
Of all sensor performance parameters, the conversion gain is arguably the most fundamental as it describes the conversion of photoelectrons at the sensor input into digital numbers at the output.  Due in part to the emergence of deep sub-electron read noise image sensors in recent years, the literature has seen a resurgence of papers detailing methods for estimating conversion gain in both the sub-electron and multi-electron read noise regimes. Each of the proposed methods work from identical noise models but nevertheless yield diverse procedures for estimating conversion gain. Here, an overview of the proposed methods is provided along with an investigation into their assumptions, uncertainty, and measurement requirements. A sensitivity analysis is conducted using synthetic data for a variety of different sensor configurations.  Specifically, the dependence of the conversion gain estimate uncertainty on the magnitude of read noise and quanta exposure is explored. Guidance into the trade-offs between the different methods is provided so that experimenters understand which method is optimal for their application. In support of the reproducible research effort, the MATLAB functions associated with this work can be found on the Mathworks file exchange.
\end{abstract}


%%%%%%%%%%%%%%%%%%%%%%%%%%%%%%%%%%%%%%%%%%%%%%%%%%%%%%%%%%%%%%%%
%%%%%%%%%%%%%%%%%%%%%%%%%%%%%%%%%%%%%%%%%%%%%%%%%%%%%%%%%%%%%%%%
%%%%%%%%%%%%%%%%%%%%%%%%%%%%%%%%%%%%%%%%%%%%%%%%%%%%%%%%%%%%%%%%

\section{INTRODUCTION}
\label{sec:intro}  % \label{} allows reference to this section

Since the advent of the Charge-Coupled Device (CCD) in the early 1970s, methods for characterizing electro-optical image sensors have continued to adapt to emerging technologies.  Of particular importance in image sensor characterization is the measurement of conversion gain, which describes an intrinsic conversion constant relating arbitrary units of Digital Numbers (DN) at the sensor output back to a physically meaningful quantity of electrons ($e\text{-}$) at the sensor input.  Generally speaking, each pixel in an image sensor array will have a unique conversion gain and this gain nonuniformity corrupts the output imagery.  For this reason, a precise estimate of each pixel's conversion gain is needed to correct the image degrading effects of gain nonuniformity and calibrate the sensor in terms of absolute units. For many decades, the Photon Transfer (PT) method has been the standard approach to conversion gain estimation\cite{Beecken:96,janesick_2001,janesick_2007,EMVA_1288_4_linear}.  Since the arrival of photons at a sensor is accurately modeled by the Poisson distribution, the moments of the sensor input are known allowing PT to treat the sensor as a black box, only observing the statistical moments of the output, to determine the conversion gain.

\begin{figure}[htb]
    \centering
    \includegraphics[scale=0.6]{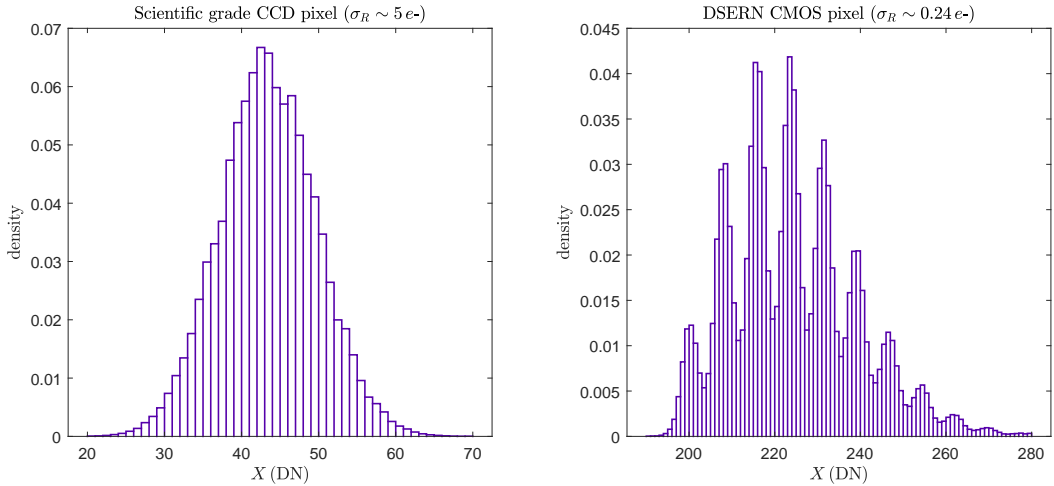}
    \caption{Histograms produced by a scientific grade CCD pixel (left) and DSERN CMOS pixel (right).}
    \label{fig:camera_comparison}
\end{figure}

In 2015, the first Deep Sub-Electron Read Noise (DSERN) image sensor was reported in the literature, carrying with it promising applications in low-light imaging and quantum technologies \cite{fossum_2015}.  As a result of sub-electron read noise, \emph{DSERN} devices could \emph{discern} the number of electrons generated in each pixel leading to never before seen structure in the data produced by such devices. As an example, Figure \ref{fig:camera_comparison} shows histograms produced by a traditional scientific grade CCD pixel (left) and DSERN CMOS pixel (right) exposed to constant illumination.  While it is not possible to observe electron events in the CCD data (since the read noise $\sigma_R$ is too large), the DSERN produced histogram shows distinct peaks where zero, one, two, etc. free-electrons have been detected within the pixel.

The additional structure observed in DSERN sensor data has led to the development of several new methods of conversion gain estimation, which leverage the additional structure to produce lower uncertainty estimates in comparison to the traditional PT method\cite{starkey_2016,Nakamoto_2022,hendrickson_2023}. What is not clear in this body of research is that all of the proposed methods are derived from the same statistical model, which is valid for sensors with sub-electron and multi-electron read noise.  Furthermore, while these newly proposed methods were designed to leverage the additional structure in data produced by DSERN capable devices, some show promise in characterizing sensors outside the DSERN regime; thus, serving as a general estimation procedure to supersede the legacy PT method.

In this work, a comprehensive overview of all currently available methods for conversion gain estimation will be discussed using a unified model and notation to facilitate comparison between each method. This will be accomplished by first describing the unifying model of sensor noise and then using the framework of this model to describe each method in detail.  With a full description of each method at hand, Monte Carlo simulations will be carried out to determine which method is best under a variety of sensor parameters. The authors aim to implement each method as faithfully as possible, and to this end, all of the code, including the implementation of each method, is available on the Mathworks file exchange.

%%%%%%%%%%%%%%%%%%%%%%%%%%%%%%%%%%%%%%%%%%%%%%%%%%%%%%%%%%%%%%%%
%%%%%%%%%%%%%%%%%%%%%%%%%%%%%%%%%%%%%%%%%%%%%%%%%%%%%%%%%%%%%%%%
%%%%%%%%%%%%%%%%%%%%%%%%%%%%%%%%%%%%%%%%%%%%%%%%%%%%%%%%%%%%%%%%

\section{The Photon Counting Distribution Model}
\label{sec:PCD_model}

The Photon Counting Distribution (PCD) model represents a single observation from a pixel (a digital gray value) as the random variable $X$ given by\cite{hendrickson_2023,hendrickson_2023_2}
\begin{equation}
\begin{aligned}
    \label{eq:PCD_RV_definition}
    X &=\lceil (K+R)/g+\mu\rfloor\\
    K &\sim\operatorname{Poisson}(H)\\
    R &\sim\mathcal N(0,\sigma_R^2).
\end{aligned}
\end{equation}
The variables used in this model are defined as follows: $K$ represents the \emph{electron number}, $H$ represents the expected number of electrons generated (thermally or otherwise) per integration time and is expressed in units of ($e\text{-}$), $\sigma_R$ represents the input referred analog read noise in units of ($e\text{-}$), $g$ represents the conversion gain in units of ($e\text{-}/\mathrm{DN}$), $\mu$ represents the pixel bias or DC offset in units of (DN), and $\lceil\cdot\rfloor$ denotes rounding to the nearest integer. In all, the random variable $X$ captures the process of adding noise $(R)$ to a number of electrons $(K)$ followed by the application of gain, offset, and finally quantization.

The quantization (rounding) defining the PCD model in (\ref{eq:PCD_RV_definition}) adds significant complexity to the distribution of $X$. If, however, $g\ll\sigma_R$ so that the quantization bins are sufficiently small, the quantization process can be modeled as an additive noise source so that a continuous distribution still provides an adequate model. Under this assumption the distribution of $X$ is modeled by the PCD
\begin{equation}
\label{eq:PCD_series_form}
f_X(x|\theta)=\sum_{k=0}^\infty\frac{e^{-H}H^k}{k!}\phi(x;\mu+k/g,\sigma^2),
\end{equation}
where $\theta=(H,g,a,b^2)$ denotes the PCD parameter vector, $\phi(x;\alpha,\beta^2)$ is the Gaussian probability density with mean $\alpha$ and variance $\beta^2$, and $\sigma=(\sigma_R^2/g^2+\sigma_Q^2)^{1/2}$ represents the combined read and quantization noise in units of (DN). For most applications the series representation (\ref{eq:PCD_series_form}) works best since only a few terms are needed to get a good approximation for $f_X$; however, through the use of characteristic functions an integral representation can also be derived in the form
\begin{equation}
\label{eq:PCD_integral_form}
f_X(x|\theta)=\frac{1}{\pi}\int_0^\infty\exp(H(\cos(t/g)-1)-\sigma^2t^2/2)\cos((\mu-x)t+H\sin(t/g))\,\mathrm dt.
\end{equation}
Furthermore, (\ref{eq:PCD_series_form}) suggests a Monte Carlo estimator of the form
\begin{equation}
    f_X(x|\theta)=\mathsf E\phi(x;\mu+K/g,\sigma^2)\approx\frac{1}{n}\sum_{k=1}^n\phi(x;\mu+K_k/g,\sigma^2),
\end{equation}
where $\{K_1,\dots,K_n\}$ are i.i.d.~$\operatorname{Poisson}(H)$ random variables. For notational purposes the shorthand $X\sim\operatorname{PCD}(H,g,\mu,\sigma^2)$ will be used to denote a random variable distributed according to the PCD. Two special cases of the PCD occur as $H\to 0$ and $\sigma\to\infty$ giving $\operatorname{PCD}(H,g,\mu,\sigma^2)\to\mathcal N(\mu,\sigma^2)$ and $\operatorname{PCD}(H,g,\mu,\sigma^2)\to\mathcal N(\mu+H/g,\sigma^2+H/g^2)$, respectively.

\begin{figure}[htb]
    \centering
    \includegraphics[scale=0.9]{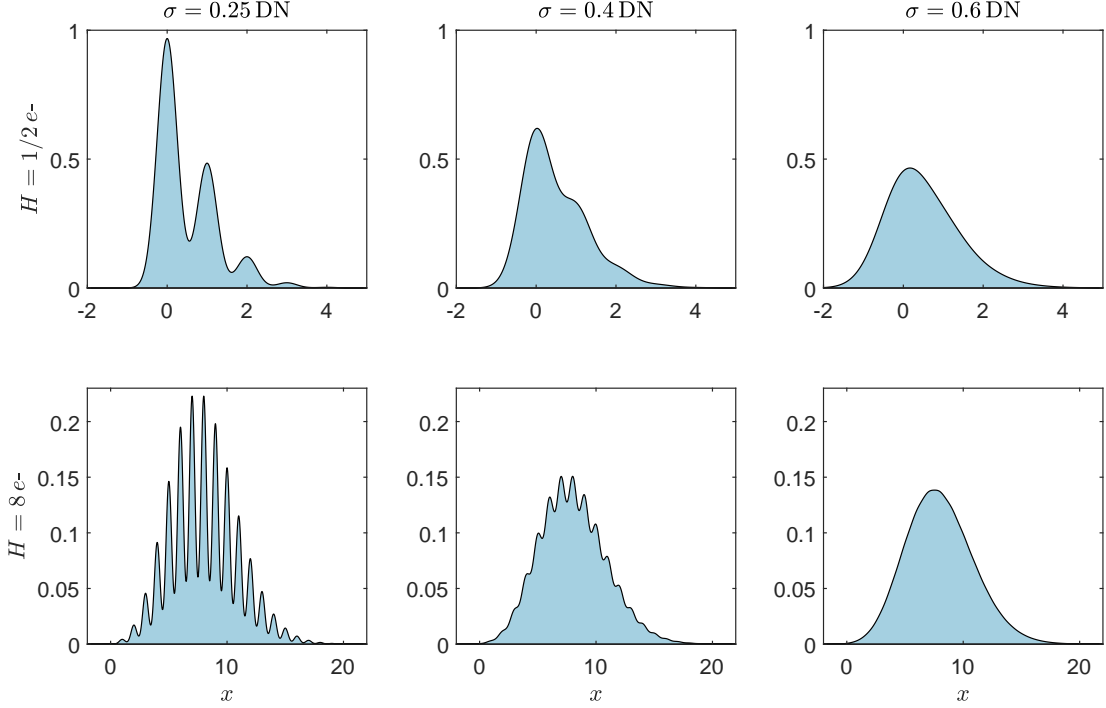}
    \caption{Plots of the PCD for various $H$ and $\sigma^2$ with $\mu=0$ and $g=1$ fixed.}
    \label{fig:PCD_plots}
\end{figure}

As far as the author's know, the first known mention of the PCD (albeit not by this name), in the context of image sensors, can be found is James Janesick's \emph{Photon Transfer} (pg.~26, Figure 3.7), which showed simulated data for the standardized version $\operatorname{PCD}(1,1,0,\sigma^2)$ \cite{janesick_2007}. Later papers by Fossum, Starkey, and Ma \cite{fossum_2013,starkey_2016,fossum_2016,Ma_2017,fossum_2022} wrote down a more complete mathematical description of $\operatorname{PCD}(H,1,0,\sigma^2)$, which included a parameter for the quanta exposure.  Furthermore, Nakamoto and Hotaka\cite{Nakamoto_2022} included a parameter for the gain in the form $\operatorname{PCD}(H,g,0,\sigma^2)$ but never in the full form accounting for the offset as seen in (\ref{eq:PCD_series_form}). What makes (\ref{eq:PCD_series_form}) a complete description is the fact that it provides all the parameters necessary to fit the PCD to raw sensor data. Depending on the specified parameters, the shape of the PCD can vary from a simple Gaussian bell-curve to a more complicated form involving many local maxima (peaks).  The parameter $\mu$ acts as a location parameter shifting the PCD on the x-axis, while $g$ acts as a scaling factor that controls the distance between adjacent peaks.  Changing either of these parameters does not drastically change the overall look of the PCD. On the other hand, the parameters $H$ and $\sigma^2$ play a significant role in the shape of the PCD. Figure \ref{fig:PCD_plots} plots the PCD for various $H$ and $\sigma^2$ (fixing $\mu=0$ and $g=1$) to show how these parameters change the appearance of the probability density.  In particular, for small enough $\sigma^2$, the PCD oscillates showing many local maxima. Sensors that exhibit clearly resolved peaks like this are said to belong to the DSERN (a.k.a.~sub-electron noise) regime. Likewise, the parameter $H$ changes the overall envelope of the PCD from a highly skewed form at small $H$ to a more Gaussian profile at large $H$.

%%%%%%%%%%%%%%%%%%%%%%%%%%%%%%%%%%%%%%%%%%%%%%%%%%%%%%%%%%%%%%%%
%%%%%%%%%%%%%%%%%%%%%%%%%%%%%%%%%%%%%%%%%%%%%%%%%%%%%%%%%%%%%%%%
%%%%%%%%%%%%%%%%%%%%%%%%%%%%%%%%%%%%%%%%%%%%%%%%%%%%%%%%%%%%%%%%

\section{Methods for Estimating conversion gain}
\label{sec:methods}

In recent years several novel methods have emerged for estimating conversion gain in the sub-electron and multi-electron noise regimes. What is not immediately clear in the literature is that all of these newly proposed methods, as well as the traditional PT method, can all be fully described in the context of the PCD model introduced in the previous section. As such, the goal of this section is to explain each method, using a unified notation, in the PCD framework. Supporting theory for each method will be presented followed by a discussion of the associated pros and cons.

%%%%%%%%%%%%%%%%%%%%%%%%%%%%%%%%%%%%%%%%%%%%%%%%%%%%%%%%%%%%%%%%

\subsection{Photon Transfer Method}
\label{subsec:photon_transfer}

Photon Transfer (PT) is a classic method for measuring conversion gain that has been around for several decades in various forms \cite{janesick_2007,Beecken:96,Hendrickson:17}. To derive the PT estimator of the conversion gain, first let $X\sim\operatorname{PCD}(H,g,\mu,\sigma^2)$ and define $v(H)\coloneqq\mathsf{Var}X=\sigma^2+H/g^2$ to be the variance of $X$ as a function of the quanta exposure $H$. Here, the symbol $\mathsf E$ is used to denote the expectation operator so that the variance is defined as $\mathsf{Var}X=\mathsf EX^2-(\mathsf EX)^2$. It follows from the definition of the derivative that
\begin{equation}
\label{eq:g_derivative_form}
\frac{1/g}{\partial_H v(H)}=\lim_{\Delta H\to 0}\frac{\Delta H/g}{v(H+\Delta H)-v(H)}=\lim_{\Delta H\to 0}g=g.    
\end{equation}
Notice that the fraction defining the derivative in (\ref{eq:g_derivative_form}) is independent of $\Delta H$; thus, the limit $\Delta H\to 0$ is not needed and nonzero values of $\Delta H$ may be used to compute $g$. With this information now suppose $X\sim\operatorname{PCD}(H+\Delta H,g,\mu,\sigma^2)$ and $Y\sim\operatorname{PCD}(H,g,\mu,\sigma^2)$ are independent PCD random variables. Consequently, $\Delta H/g=\mathsf EX-\mathsf EY$ and $v(H+\Delta H)-v(H)=\mathsf{Var}X-\mathsf{Var}Y$ leading to the classic PT relation
\begin{equation}
\label{eq:g_moment_form}
    g=\frac{\mathsf EX-\mathsf EY}{\mathsf{Var}X-\mathsf{Var}Y}.
\end{equation}

The PT method for conversion gain estimation replaces the populations means and variances in (\ref{eq:g_moment_form}) with their respective unbiased estimators to obtain an estimator for $g$. Specifically, let $\mathbf x=\{x_1,\dots,x_{n_1}\}$ with $x_k\sim\operatorname{PCD}(H+\Delta H,g,\mu,\sigma^2)$ denote a random sample of $n_1$ observations at a quanta exposure of $H+\Delta H$ and $\mathbf y=\{y_1,\dots,y_{n_2}\}$ with $y_k\sim\operatorname{PCD}(H,g,\mu,\sigma^2)$ denote a second (independent) random sample of $n_2$ observations at a quanta exposure of $H$. Denoting $\bar x=\frac{1}{n_1}\sum_{k=1}^{n_1}x_k$ as the sample mean and $\hat x=\frac{1}{n_1-1}\sum_{k=1}^{n_1}(x_k-\bar x)^2$ as the sample variance of the $\mathbf x$-sample (and likewise for $\bar y$ and $\hat y$), the PT estimate for the conversion gain is given by \cite{Hendrickson:17,hendrickson_22}
\begin{equation}
    \tilde g=\frac{\bar x-\bar y}{\hat x-\hat y}.
\end{equation}

Since this is an estimator of two independent samples, Hendrickson et.~al.~(2022) derived approximate optimal sample size pairs $(n_1^\mathrm{opt},n_2^\mathrm{opt})$ of the form \cite{hendrickson_22}
\begin{equation}
    \label{eq:optimal_sample_sizes}
    \begin{aligned}
        n_1^\mathrm{opt} &\sim\frac{2(1+\zeta)}{\mathsf{acv}_0^2(1-\zeta)^2}+5\\
        n_2^\mathrm{opt} &\sim\frac{2\zeta(1+\zeta)}{\mathsf{acv}_0^2(1-\zeta)^2}+1,
    \end{aligned}
\end{equation}
where
\begin{equation}
\zeta=\frac{\mathsf{Var}Y}{\mathsf{Var}X}=\frac{\sigma^2+\frac{H}{g^2}}{\sigma^2+\frac{H}{g^2}+\frac{\Delta H}{g^2}}
\end{equation}
and $\mathsf{acv}_0$ denotes the desired relative uncertainty of the final estimate, e.g.~$\mathsf{acv}_0=0.05$ corresponds to $5\%$ estimator uncertainty \cite{hendrickson_22}. These approximate optimal sample size pairs allow an experimenter to achieve the desired estimate uncertainty ($\mathsf{acv}_0$) with the fewest total number of samples possible \cite{hendrickson_22}.

In practice, most image sensors are not perfectly linear ($g$ is dependent on $H$) so that a small $\Delta H$ ($\zeta\approx 1$) is needed to obtain a meaningful estimate of $g$ at the chosen illumination level. This, however, can cause instability in the estimator due to the fact that statistical uncertainty in the quantity $\hat x-\hat y$ can lead to division by zero type errors\footnote{The numerical instability of this estimator for $g$ is similar to the numerical instability of numerical derivatives.}. In fact, for even moderately large $n_1$ and $n_2$, the sampling distribution of $(\hat x-\hat y)^{-1}$ is accurately modeled by the inverse gamma-difference distribution, which is known to have undefined moments in a similar manner as the Cauchy distribution \cite{hendrickson_2019}. This lack of well-defined moments leads to the PT estimator $\tilde g$ having ill-behaved statistical properties, most of which can be mitigated by using very large sample sizes (notice that $n_i^\mathrm{opt}\to\infty$ and $\Delta H\to 0$). Additionally, a disadvantage of this estimator is that it utilizes only the first two moments of the PCD.  Since the PCD is not fully described by these first two moments, the PT estimator does not fully utilize all the information about $g$ contained in the sample leading to larger estimator uncertainty compared to other techniques. Despite these disadvantages, the PT estimator is still attractive as it provides useful estimates of $g$ in both the sub-electron and multi-electron read noise regimes and is calculated from basic sample moments; rendering it the most computationally inexpensive estimator of $g$.

%%%%%%%%%%%%%%%%%%%%%%%%%%%%%%%%%%%%%%%%%%%%%%%%%%%%%%%%%%%%%%%%

\subsection{Photon Counting Histogram Method}
\label{subsec:PCH_method}

\begin{figure}[htb]
    \centering
    \includegraphics[scale=0.65]{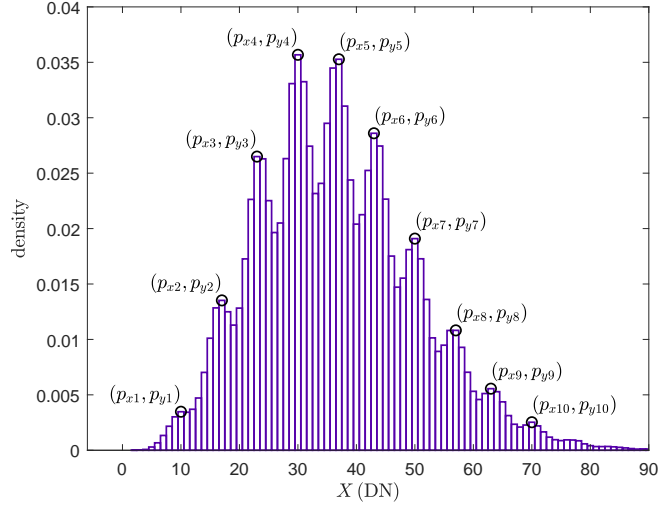}
    \caption{Simulated PCH showing ten detected peak locations.}
    \label{fig:PCH_detected_peaks}
\end{figure}

In response to the emergence of DSERN capable image sensors, the Photon Counting Histogram (PCH) method, developed at Dartmouth University, was the first documented method to explicitly incorporate the PCD model into the estimation of the sensor performance parameters \cite{fossum_2015,starkey_2016}. PCH is primarily a method for estimating conversion gain and read noise by detecting the locations of local maxima and minima observed in an experimentally generated histogram. To perform PCH characterization, a sample $\mathbf x=\{x_1,\dots,x_{n_1}\}$ with $x_k\sim\operatorname{PCD}(H,g,\mu,\sigma^2)$ is captured and binned as a histogram, which is the experimental PCH. Each bin count is divided by the sample size $n_1$ to normalize the histogram so that it represents an approximation of the pixel's PCD at the chosen value of $H$. Assuming the read noise is small enough, many peaks (local maxima) in the experimental PCH should be present and an algorithm for detecting these peaks is deployed. Figure \ref{fig:PCH_detected_peaks} shows a simulated PCH with the locations of ten detected peaks.

To estimate the conversion gain, let $\{(p_{x1},p_{y1}),\dots,(p_{xm},p_{ym})\}$ denote a sequence of $m$ consecutive peak locations detected in the experimental PCH.  According to the PCD model, the abscissas of the peaks locations $\{p_{xk}\}$, in units of $(\mathrm{DN})$, are approximately located at equally spaced intervals of the form $p_{xk}=\mu+k/g$ with $k\in\Bbb N_0$. As such, fitting a line to the data $\{(1,p_{x1}),\dots,(m,p_{xm})\}$ and extracting the reciprocal slope of the fit yields an estimate $\tilde g$ for the conversion gain. If the electron number associated with each peak location is also known, one may instead fit a line to the data $\{(k_1,p_{x1}),\dots,(k_m,p_{xm})\}$ with the reciprocal slope  again yielding $\tilde g$ and the $y$-intercept yielding an estimate for the bias $\tilde\mu$.

\begin{figure}[htb]
    \centering
    \includegraphics[scale=0.65]{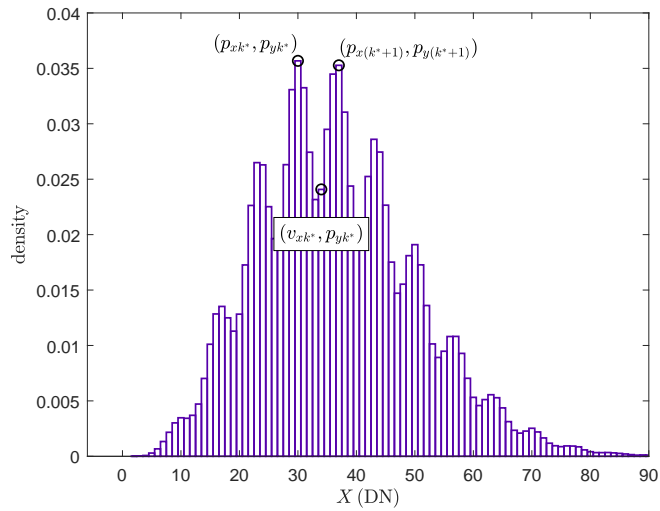}
    \caption{Simulated PCH showing two most prominent peaks with corresponding valley location used for quanta exposure and read noise estimation.}
    \label{fig:PCH_detected_peaks2}
\end{figure}

Estimation of the quanta exposure requires the two most prominent peak locations and their electron numbers denoted by $\{(p_{xk^\ast},p_{yk^\ast}),(p_{x(k^\ast+1)},p_{y(k^\ast+1)})\}$ and $(k^\ast,k^\ast+1)$, respectively (see Figure \ref{fig:PCH_detected_peaks2}). For small read noise values the ordinates of these peaks are approximated by
\begin{equation}
    p_{yk}\sim\frac{1}{\sqrt{2\pi}\sigma}\frac{e^{-H}H^k}{k!}.
\end{equation}
Taking the ratio of the two most prominent peaks and solving for $H$ subsequently gives the estimate (see Figure \ref{fig:PCH_detected_peaks2})
\begin{equation}
    \tilde H=(k^\ast+1)\frac{p_{y(k^\ast+1)}}{p_{yk^\ast}}.
\end{equation}
In a similar manner, the read noise is calculated by first locating the valley (local minima) between the two most prominent peaks, denoted $(v_{x\ast},v_{y\ast})$, and then computing the Valley Peak Modulation (VPM)
\begin{equation}
    \operatorname{VPM}=1-\frac{v_{y\ast}}{\frac{1}{2}(p_{yk^\ast}+p_{y(k^\ast+1)})}.
\end{equation}
The VPM is independent of the parameters $g$ and $\mu$ so that a lookup table can be generated containing the VPM for various values of $\sigma_R$ and $H$. Using the estimate $\tilde H$, one can then lookup the value of $\sigma_R$ corresponding to the estimated VPM to obtain the PCH estimate of read noise.

Assuming one is able to obtain estimates for all four parameters (which requires knowing the electron numbers for the detected peaks), the PCH estimates can be refined by fitting the PCD to the experimental PCH using nonlinear least squares with the initial parameter estimates as starting points.

PCH provides an intuitive graphical approach to sensor characterization and only requires a single sample of data. Because PCH incorporates the full description of the PCD model into the estimation procedure, it leverages the structure of DSERN data resulting in estimates of conversion gain with less uncertainty compared to the PT method. Furthermore, a unique feature of PCH is that the initial estimate of $g$ obtained from peak locations does not assume a Poissonian light source so that sources of an arbitrary probabilistic nature can in theory be used\cite{Nakamoto_2022}. The biggest challenge with this method is the need to reliably detect peak and valley locations, which requires both large sample sizes and sufficiently small read noise so that the peaks can be observed. For this reason, the applicability of the PCH method is restricted to the DSERN regime.

%%%%%%%%%%%%%%%%%%%%%%%%%%%%%%%%%%%%%%%%%%%%%%%%%%%%%%%%%%%%%%%%

\subsection{Fourier Transform Method}
\label{subsec:fft_method}

Initially, the Fourier-based approach for sensor characterization was not developed as a self-contained method, but rather as a means of deriving starting points for the PCH-EM algorithm discussed in Section \ref{subsec:PCHEM_method}. Nevertheless, this approach can be regarded as a characterization method in its own right\cite{hendrickson_2023}. The idea behind this technique, similar to the PCH method, stemmed from the fact that the PCD exhibits periodic oscillations when the read noise is sufficiently small. This approach revolves around the analytical expression for the magnitude of the PCD Fourier transform given by
\begin{equation}
    \label{eq:FT_magnitude}
    |\hat f_X(\omega)|\coloneqq|\mathsf E\exp(-2\pi i\omega X)|=\exp(H(\cos(2\pi\omega/g)-1)-2\pi^2\sigma^2\omega^2).
\end{equation}

\begin{figure}[htb]
    \centering
    \includegraphics[scale=0.65]{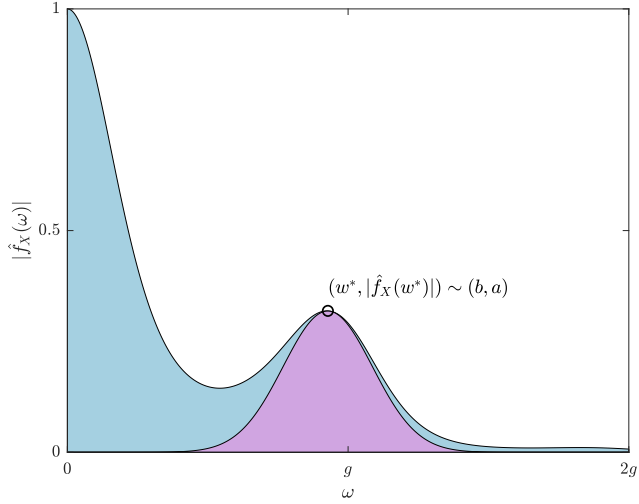}
    \caption{Graph of $|\hat f_X(\omega)|$ (blue) and its asymptotic approximation near $\omega=g$ (purple) versus $\omega$ showing the two most dominant peaks at $\omega=0$ and $\omega=\omega^\ast$.}
    \label{fig:FT_mag}
\end{figure}

Recall that when the read noise is sufficiently small, the PCD has local maxima occurring with a period of approximately $1/g$ (frequency of $g$). This property means the magnitude function should exhibit local maxima at frequencies approximately located at integer multiple of $g$ as is seen in Figure \ref{fig:FT_mag} (blue curve). Let $\omega^\ast$ denote the frequency corresponding to the local maxima near $g$, which is the second most prominent peak of the magnitude function after the primary peak at $\omega=0$. We note that in order for this secondary peak to exist we must have $\sigma_{e\text{-}}^2/H<|\min_{x>0}\operatorname{sinc}x|=0.217\dots$, where $\sigma_{e\text{-}}=\sigma\times g$ is the read plus quantization noise in units of electrons and $\operatorname{sinc}x=\sin x/x$. Using Lagrange inversion and defining $z=-\sigma_{e\text{-}}^2/H$ we compute
\begin{equation}
\label{eq:Lagrange_inversion}
    \omega^\ast=g+\sum_{n=1}^\infty\lim_{\omega\to g} \partial_\omega^{n-1}\left(\frac{\omega-g}{\operatorname{sinc}(2\pi\omega/g)}\right)^n\frac{z^n}{n!} =g(1+z+z^2+\mathcal O(z^3)),
\end{equation}
which shows that for small $|z|$, $\omega^\ast$ is very well approximated by $g$.

With this information we can approximate the magnitude function $|\hat f_X(\omega)|$ near the secondary peak by considering the following asymptotic approximation as $\omega\to g$:
\begin{equation}
    \label{eq:FT_mag_1st_peak}
    |\hat f_X(\omega)|\sim a\exp(-2\pi^2v(\omega-b)^2),
\end{equation}
where $v=\mathsf{Var}X=\sigma^2+H/g^2$,
\begin{equation}
    a=\exp\left(-2\pi^2\left(H-\frac{(H/g)^2}{v}\right)\right),   
\end{equation}
and
\begin{equation}
    b=\frac{H/g}{v}.    
\end{equation}
Figure \ref{fig:FT_mag} shows the exact magnitude function (blue) compared to the asymptotic approximation (purple) along with the location of the secondary peak $(\omega^\ast,|\hat f_X(\omega^\ast)|)$. As can be observed, the peak of the asymptotic approximation, $(b,a)$, provides an excellent approximation to the location of the exact peak. To understand why this is, notice that this asymptotic expression approximates $\omega^\ast$ as
\begin{equation}
    \omega^\ast\sim b=\frac{g}{1+\sigma_{e\text{-}}^2/H}=\frac{g}{1-z}=g(1+z+z^2+\mathcal O(z^3)),
\end{equation}
which shows agreement with the first three terms of the exact expansion for $\omega^\ast$ obtained in (\ref{eq:Lagrange_inversion}).

The system of three equations given by $v$, $a$, and $b$ can be inverted to give the following approximations for $H$, $g$, and $\sigma^2$:
\begin{equation}
\begin{aligned}
    \label{eq:parameter_approximations}
    H(v,a,b) &\sim vb^2-\frac{\log a}{2\pi^2}\\
    g(v,a,b) &\sim b-\frac{\log a}{2\pi^2vb}\\
    \sigma^2(v,a,b) &\sim v-\left(v-\frac{\log a}{2\pi^2b^2}\right)^{-1}.
\end{aligned}
\end{equation}

Equipped with these details, the Fourier based method of characterization is as follows. First, a sample $\mathbf x=\{x_1,\dots,x_{n_1}\}$ with $x_k\sim\operatorname{PCD}(H,g,\mu,\sigma^2)$ is captured and the sample mean $\bar x=\frac{1}{n_1}\sum_{k=1}^{n_1}x_k$ and sample variance $\hat x=\frac{1}{n_1-1}\sum_{k=1}^{n_1}(x_k-\bar x)^2$ are computed. Since the sample $\mathbf x$ is integer-valued, it is binned in bins centered on the integers and the counts $c_k$ for each bin location $b_k$ are normalized via $p_k=c_k/n_1$ so that we obtain a density normalized experimental PCH. The Discrete Fourier Transform (DFT) of the density normalized experimental PCH is then calculated and the location of the secondary peak in the DFT is detected yielding an estimate $(\tilde b,\tilde a)$.  Initial estimates for $H$, $g$ and $\sigma^2$ are then found from (\ref{eq:parameter_approximations}) giving $(H_0,g_0,\sigma_0^2)=(H(\hat x,\tilde a,\tilde b),g(\hat x,\tilde a,\tilde b),\sigma^2(\hat x,\tilde a,\tilde b))$. Final estimates $\tilde H$, $\tilde g$, $\tilde\sigma^2$ are then found by fitting the magnitude function (\ref{eq:FT_magnitude}) to the experimental PCH DFT using nonlinear least squares and the starting points $(H_0,g_0,\sigma_0^2)$. Using the fact that $\mathsf EX=\mu+H/g$ we then obtain an estimate for $\mu$ in the form $\tilde\mu=\bar x-\tilde H/\tilde g$. Further improvements on $\tilde\mu$ are possible using autocorrelation\cite{hendrickson_2023}.

The Fourier based method echos that of the PCH method in that it requires sufficiently small read noise to work. It also requires some implementation of peak detection and refines the initial estimates with nonlinear least squares. In fact, the PCH and Fourier methods are essentially the same method with PCH operating in the original data space and Fourier operating in the frequency space. What makes the Fourier method attractive as that it obtains estimates of each PCD parameter from a single sample and requires only detecting a single peak in the experimental PCH DFT (compare this to detecting many peaks with PCH). Since only a single peak is needed and the method is based on the DFT it is also easily automated and computationally inexpensive. Additionally, this method makes full use of the PCD model so that it can usually obtain lower uncertainty estimates of the conversion gain in comparison to the PT method.  The major downside of this method is that the read noise needs to be small enough to guarantee the existence of the secondary peak. This ultimately limits the applicability of this method, like the PCH method, to the DSERN regime.

%%%%%%%%%%%%%%%%%%%%%%%%%%%%%%%%%%%%%%%%%%%%%%%%%%%%%%%%%%%%%%%%

\subsection{Nakamoto's Method}
\label{subsec:nakamoto_method}

In response to the PCH method, Nakamoto and Hotaka introduced a characterization technique based on the principle of Maximum Likelihood Estimation (MLE) that also takes advantage of the full PCD model \cite{Nakamoto_2022}. To understand how this method works, let $\mathbf y=\{y_1,\dots,y_{n_2}\}$ with $y_k\sim\operatorname{PCD}(0,g,\mu,\sigma^2)\overset{d}{=}\mathcal N(\mu,\sigma^2)$ be a sample of data captured under dark conditions with short enough integration time so that dark current is negligible $(H=0)$. Since the distribution of this data is normal, unbiased estimates for $\mu$ and $\sigma^2$ may be directly measured from the $\mathbf y$-sample via
\begin{equation}
    \tilde\mu=\frac{1}{n_2}\sum_{k=1}^{n_2}y_k
\end{equation}
and
\begin{equation}
    \tilde\sigma^2=\frac{1}{n_2-1}\sum_{k=1}^{n_2}(y_k-\tilde\mu)^2,
\end{equation}
which are the sample mean and sample variance, respectively. Now consider capturing a second sample $\mathbf x=\{x_1,\dots,x_{n_1}\}$ with $x_k\sim\operatorname{PCD}(H,g,\mu,\sigma^2)$ for some choice of $H>0$. Since $\mathsf Ex_k=\mu+H/g$ we can estimate the sample mean $\bar x=\frac{1}{n_1}\sum_{k=1}^{n_1}x_k$ and then construct an estimator for the quanta exposure as a function of $g$ in the form
\begin{equation}
    \tilde H(g)=g(\bar x-\tilde\mu).
\end{equation}
Then, a constrained likelihood function is made from these three estimates via
\begin{equation}
    L(g|\mathbf x)=\prod_{k=1}^{n_1}f_X(x_k|\tilde H(g),g,\tilde\mu,\tilde\sigma^2)
\end{equation}
and an estimate for $g$ is computed by maximizing this constrained likelihood function (or equivalently it's logarithm)
\begin{equation}
    \tilde g=\argmax_g L(g|\mathbf x).
\end{equation}
An estimate for the quanta exposure follows by evaluating $\tilde H=\tilde H(\tilde g)$. Since closed-form solutions for this maximization are intractable, any number of numerical methods can be employed.

Like the PCH and Fourier methods, Nakamoto's method incorporates the full description of the PCD model into the estimation procedure, which generally results in conversion gain estimates with less uncertainty than can be obtained with the traditional PT method. Furthermore, the $H=0$ sample allows for direct measurement of $\mu$ and $\sigma^2$, which generally helps stabilize the conversion gain estimates when the read noise is large. Because peak detection is not part of the estimation procedure, Nakamoto's method shows promise in being a viable method for both the sub-electron and multi-electron read noise regimes. The most significant disadvantage of this method is the requirement to obtain a sample at $H=0$, which may not be possible depending on the available integration times of the sensor and the magnitude of dark current. This requirement is particularly challenging when trying to characterize an entire sensor array, which will inevitably contain hot pixels that cannot achieve a quanta exposure near zero. Lastly, we note that this method requires two samples but does not make full use of the information contained in both samples. This can be seen by the fact that the $\mathbf x$-sample contains information about $\mu$ and $\sigma$; however, these parameters are estimated only from the $\mathbf y$-sample.% Lastly, we note that the maximization of the function $L(g|\mathbf x)$ requires numerical optimization, which requires step size control to ensure convergence.

%%%%%%%%%%%%%%%%%%%%%%%%%%%%%%%%%%%%%%%%%%%%%%%%%%%%%%%%%%%%%%%%

\subsection{PCH Expectation Maximization Algorithm}
\label{subsec:PCHEM_method}

The Photon Counting Histogram Expectation Maximization (PCH-EM) algorithm (in review at the time of writing) is the latest iteration of methods for performing sensor characterization based on the PCD model\cite{hendrickson_2023,hendrickson_2023_2,PCHEM_code}. It is the first technique devised to compute simultaneous maximum likelihood estimates for all four PCD parameters using only a single sample of data. This method was inspired from the fact that when the electron numbers associated with each observation are known, that is, we have the \emph{complete data} $(\mathbf x,\mathbf k)=\{(x_1,k_1),\dots,(x_{n_1},k_{n_1})\}$, closed-form maximum likelihood estimators for each PCD parameter are easily derived\cite{hendrickson_2023}. While the electron numbers cannot be directly observed, this fact motivates a latent (hidden) variables model of estimation, which is what the general expectation maximization algorithm provides. As such, PCH-EM is a specific implementation of the general EM algorithm with the PCD being the underlying distribution to be estimated.

To perform PCH-EM, a random sample $\mathbf x=\{x_1,\dots,x_{n_1}\}$ with $x_k\overset{\mathrm{iid}}{\sim}\operatorname{PCD}(H,g,\mu,\sigma^2)$ is captured. Given an initial estimate of the parameters $\theta_0=(H_0,g_0,\mu_0,\sigma_0^2)$ (obtained by one of the other methods, e.g.~Fourier or PT method), the PCH-EM algorithm iteratively updates the parameter estimates via the update equations \cite{hendrickson_2023}
\begin{subequations}\label{eq:update_eqns}
\begin{align}
H_{t+1} &=A_t \label{eq:H_update}\\
g_{t+1}&=\frac{B_t-H_{t+1}^2}{C_t-\bar xH_{t+1}} \label{eq:g_update}\\
\mu_{t+1}&=\bar x-\frac{H_{t+1}}{g_{t+1}} \label{eq:mu_update}\\
\sigma_{t+1}^2 &= \hat x-\frac{B_t-H_{t+1}^2}{g_{t+1}^2},\label{eq:sigma_update}
\end{align}
\end{subequations}
where $\bar x=\frac{1}{n_1}\sum_{k=1}^{n_1}x_k$ is the sample mean, $\hat x=\frac{1}{n_1}\sum_{k=1}^{n_1}(x_k-\bar x)^2$ is the sample variance, and
\begin{subequations}\label{eq:update_matrices}
\begin{align}
A_t &=\frac{1}{N}\sum_{n=1}^N\sum_{k=0}^\infty\gamma_{nk}^{(t)}k \label{eq:A_t}\\
B_t &=\frac{1}{N}\sum_{n=1}^N\sum_{k=0}^\infty\gamma_{nk}^{(t)}k^2 \label{eq:B_t}\\
C_t &=\frac{1}{N}\sum_{n=1}^Nx_n\sum_{k=0}^\infty\gamma_{nk}^{(t)}k, \label{eq:C_t}
\end{align}
\end{subequations}
where
\begin{equation}
    \label{eq:gamma_nk}
    \gamma_{nk}^{(t)}=\frac{\frac{e^{-H_t}H_t^k}{k!}\phi(x_n;\mu_t+k/g_t,\sigma_t^2)}{\sum_{\ell=0}^\infty\frac{e^{-H_t}H_t^\ell}{\ell!}\phi(x_n;\mu_t+\ell/g_t,\sigma_t^2)}.
\end{equation}
The $\gamma_{nk}^{(t)}$ are called \emph{membership probabilities} because they represent the probability of $x_n$ belonging to the $k$th Gaussian component of the PCD given the current parameter estimates $\theta_t$. Since PCH-EM is just a specific implementation of the more general Expectation Maximization (EM) algorithm, each iteration of the algorithm guarantees an increase in the likelihood of the sample. The algorithm halts when a specified convergence criteria is achieved.  In practical implementation, all of the series in the update equations can be truncated to finite sums by only considering the terms $k\in\{F^{-1}(\epsilon),\dots,F^{-1}(1-\epsilon)\}$, where $F^{-1}$ is the $\operatorname{Poisson}(H_t)$ quantile function and $\epsilon>0$ is a small positive number.

% In addition to the ability to simultaneously estimate all four PCD parameters, a unique byproduct of PCH-EM is a means to directly estimate the electron number associated with each observation in the sample by computing
% \begin{equation}
%     \tilde k_n=\argmax_k\gamma_{nk}^{(t)}.
% \end{equation}
% The estimated electron number $\tilde k_n$ represents an optimized estimate that takes into account the complex nature of the PCD to minimize estimation error.

PCH-EM provides many positive characteristics in that it provides maximum likelihood estimates of all the PCD parameters using a single sample of data, incorporates the full PCD model, and does not require numerical optimization, e.g.~Newton iteration, to maximize the sample likelihood. It is also easily automated and computationally inexpensive although not as inexpensive as traditional PT. Furthermore, because the general EM algorithm is so well studied, many extensions of PCH-EM are possible to improve the robustness of the algorithm and its estimates. The major downside of this method, which holds for all other methods excluding PT, is the requirement of starting points.  Poor starting points can result in slow convergence or convergence to a local maxima while missing the global maxima of the sample likelihood function. While PCH-EM can suffer form the issue of local maxima, extensions of the algorithm using annealing are possible\cite{Ueda_1994,Ueda_1998,Miyahara_2017}.

%%%%%%%%%%%%%%%%%%%%%%%%%%%%%%%%%%%%%%%%%%%%%%%%%%%%%%%%%%%%%%%%

\subsection{Two-Sample PCH-EM Algorithm}
\label{subsec:PCHEM_method}

Currently in development, PCH-EM2 is the two-sample generalization of PCH-EM that incorporates two samples taken at different $H$ into a single likelihood function, thus offering both advantages and disadvantages similar to the single-sample PCH-EM approach. One significant benefit of PCH-EM2 is that it enables one to obtain low-uncertainty estimates of all four parameters by combining samples taken at two different $H$-values, as the uncertainty of estimates for each PCD parameter varies differently with $H$. Therefore, PCH-EM2 is expected to be more accurate than PCH-EM. Further extensions of this method to an arbitrary number of samples is also possible. Additionally, the use of an extra sample stabilizes the algorithm in the presence of high read noise, making it a good candidate for general estimation procedure in the sub-electron and multi-electron read noise regimes. Unlike Nakamoto's method, PCH-EM2 can extract information about each parameter from both samples, making it theoretically more accurate. However, PCH-EM2, like other two-sample methods, is more sensitive to nonlinearity in comparison to single-sample methods because it requires the sensor to behave linearly over both samples instead of just one.

%%%%%%%%%%%%%%%%%%%%%%%%%%%%%%%%%%%%%%%%%%%%%%%%%%%%%%%%%%%%%%%%
%%%%%%%%%%%%%%%%%%%%%%%%%%%%%%%%%%%%%%%%%%%%%%%%%%%%%%%%%%%%%%%%
%%%%%%%%%%%%%%%%%%%%%%%%%%%%%%%%%%%%%%%%%%%%%%%%%%%%%%%%%%%%%%%%

\section{Comparison of methods}
\label{sec:methods_comparison}

\subsection{Design of Experiment}

To compare the uncertainty of each methods' conversion gain estimates, Monte Carlo experiments were performed. However, the four-dimensional parameter space of the PCD model made it challenging to fully explore in simulations. To reduce the area of exploration, the parameter $\mu$ was set to zero for all simulation runs since it only shifts the PCD without changing its shape. Moreover, to avoid over-quantization, the conversion gain was fixed at $g=\sigma_R/6$ for all runs, leaving only two dimensions $(\sigma_R,H)\in\Bbb R_{>0}\times\Bbb R_{\geq 0}$ to consider. As the read noise surpasses about $0.5\,e\text{-}$, peak detection methods struggle, so the read noise interval was limited to $\sigma_R\in(0,2)$. Similarly, the Poisson distribution's dynamic changes occur mainly when $H$ is small, so the quanta exposure was limited to $H\in(0,10)$. Using these constraints, a grid of $64$ $\sigma_R$-values on $(0.05,2)$ and $32$ $H$-values on $(0.05,10)$ was created, which paired with $\mu=0$ and $g=\sigma_R/6$, resulted in a total of $2048$ points in the PCD parameter space to simulate data on.

\begin{figure}[htb]
    \centering
    \includegraphics[scale=0.7]{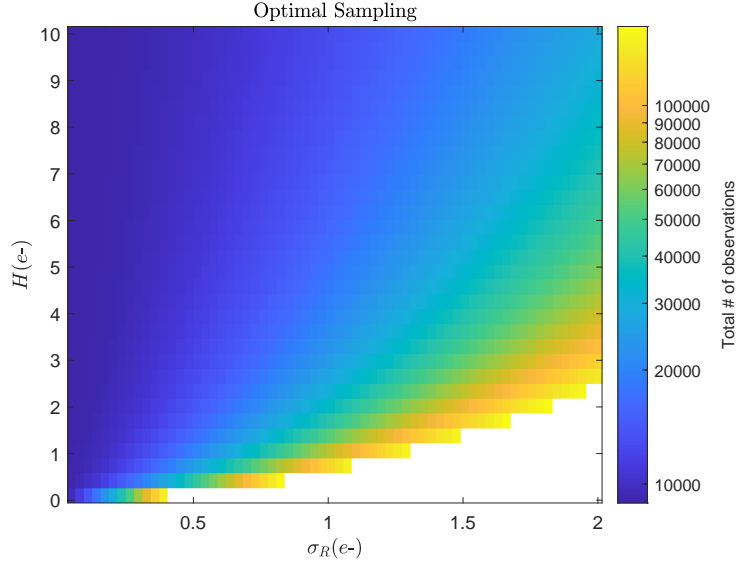}
    \caption{Total number of samples used for each set of PCD parameters. The black region corresponds to points where $n_1>10^5$, which were ignored in the simulation.}
    \label{fig:optimal_samples}
\end{figure}

Once the desired parameters were selected, the following stage in the experimental design involved selecting the kinds of data to simulate along with their corresponding sample sizes. Six methods were available, three of which (PCH, Fourier, PCH-EM) necessitated just one sample, while the remaining three (PT, Nakamoto, PCH-EM2) required two samples. Specifically, Nakamoto's approach required two samples, with one of them being at $H=0$. To accommodate all methods two types of data were generated including \emph{dark} samples of the form $\mathbf y=\{y_1,\dots,y_{n_2}\}$ with $y_k\sim\operatorname{PCD}(0,g,\mu,\sigma^2)$ and \emph{illuminated} samples of the form $\mathbf x=\{x_1,\dots,x_{n_1}\}$ with $x_k\sim\operatorname{PCD}(H,g,\mu,\sigma^2)$. Observations in each sample were generated according to the model (\ref{eq:PCD_RV_definition}). The uncertainty in any method's conversion gain estimates will be a function of the parameters. For this reason, the optimal sample size pairs in (\ref{eq:optimal_sample_sizes}) for $\mathsf{acv}_0=0.015$ and $\zeta=(1+H/\sigma_R^2)^{-1}$ were chosen to make the uncertainty in the PT conversion gain estimates mostly independent of the parameters. In this way, PT would be a reference to compare the uncertainty of the five other methods against. The specified value of $\mathsf{acv}_0=0.015$ means that the PT conversion gain estimates should have an uncertainty of approximately $1.5\%$ across all parameters. Figure \ref{fig:optimal_samples} shows the total sample size $n_1+n_2$ used as a function of the parameters. Parameters where $n_1>10^5$ (seen as the white region in Figure \ref{fig:optimal_samples}) were ignored to make sure the experiment did not take too long to complete.

To ensure a fair comparison of each method, both the dark and illuminated data were made available to all six methods, even if the method naturally used only one sample. This way, the two-sample methods did not have access to more information than the one-sample methods. The two-sample methods required no changes to their approach since they inherently incorporated both samples into their estimation procedure. However, for the one-sample methods, the information in the dark sample was integrated into the estimation procedure by providing the starting points $g_0=(\bar x-\bar y)/(\hat x-\hat y)$, $\mu_0=\bar y$, $\sigma_0^2=\hat y$, and $H_0=g_0(\bar x-\mu_0)$, where $\bar x$ represents the sample mean and $\hat x$ represents the sample variance for the $x$-data (and likewise for the $y$-data). With this step, the experimental design phase was concluded.

%%%%%%%%%%%%%%%%%%%%%%%%%%%%%%%%%%%%%%%%%%%%%%%%%%%%%%%%%%%%%%%%%%%%%%%%%%

\subsection{Results}

The experiment was executed on MATLAB code containing two nested loops.  In the outer loop, each iteration consisted of selecting the next set of PCD parameters and associated sample sizes.  For each iteration of the outer loop, the inner loop was repeated 512 times, where in each of the 512 iterations a $\mathbf x$ and $\mathbf y$ sample were generated and then supplied to each method so that the conversion gain could be estimated. This subsequently resulted in 512 conversion gain estimates $\tilde g_k$ for each method, which were then used to compute the normalized Root Mean Squared Error (RMSE)
\begin{equation}
    \operatorname{RMSE}(\theta)=\left(\frac{1}{512}\sum_{k=1}^{512}(1-\tilde g_k/g)^2\right)^{1/2}.
\end{equation}
As a result of the Monte Carlo experiment, a $64\times 32$ array of normalized RMSE values for each method was generated.

\begin{figure}[htb]
    \centering
    \includegraphics[scale=0.55]{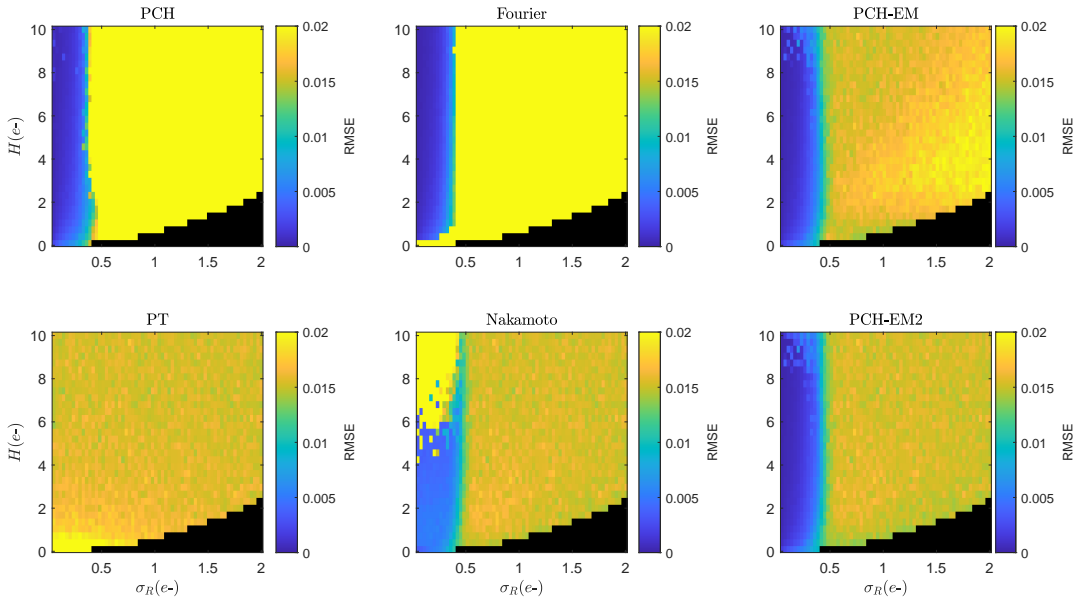}
    \caption{Comparison of six characterization methods in terms of the RMSE achieved for conversion gain estimates. Single-sample methods are shown in first row with two-sample methods in the bottom row.}
    \label{fig:All_methods_RMSE}
\end{figure}

Figure \ref{fig:All_methods_RMSE} shows the Monte Carlo estimated RMSE for each method as a function of $\sigma_R$ and $H$.  The first row is comprised of the one-sample methods with the second containing only the two-sample methods. The black region corresponds to parameters where data was not simulated due to the sample sizes becoming too large.

Several observations can be derived from the Figure \ref{fig:All_methods_RMSE}. Initially, it should be acknowledged that the PT estimates' RMSE remained relatively stable at approximately $\operatorname{RMSE}\approx 0.015$. This value corresponds to the $\mathsf{acv}_0=0.015$ uncertainty specification used for the optimal sample sizes. Therefore, the optimal sample sizes effectively controlled the PT estimate uncertainty for the parameters considered. Additionally, all five methods that utilized the full description of the PCD in the estimation process showed a region below $\sigma_R\approx 0.42,e\text{-}$ where their conversion gain estimates' uncertainty was generally lower than PT (blue strip). This finding was due to the PCD-inclusive techniques utilizing the data's extra structure at low read noise values, which PT ignores by only incorporating the first two moments.

% A number of observations can be made about the contents of Figure \ref{fig:All_methods_RMSE}.  First, it should be noted that the RMSE of the PT estimates remained more or less constant around a value of $\operatorname{RMSE}\approx 0.015$, which is equal to the uncertainty specification $\mathsf{acv}_0=0.015$ baked into the optimal sample sizes. This shows the optimal sample sizes did a good job of controlling the PT estimate uncertainty for the parameters considered. Second, all of the remaining five methods that incorporated the full form of the PCD into the estimation procedure show a region below $\sigma_R\approx 0.42\,e\text{-}$ where the uncertainty in their conversion gain estimates is generally less than those of PT (blue strip).  This feature is observed because methods utilizing the PCD leverage the additional structure in the data exhibited at low read noise values while PT ignores this extra structure by only using information about the first two moments.

Regarding the one-sample methods, the PCH and Fourier techniques outperformed PT in terms of RMSE below approximately $\sigma_R\approx 0.42,e\text{-}$; however, their performance degraded above this read noise value due to their reliance on peak detection. Unlike the PCH and Fourier methods, PCH-EM did not necessitate observing peaks and could still estimate the conversion gain at higher read noise values, bridging the gap into the multi-electron read noise regime for one-sample methods. For the two-sample methods, both Nakamoto's method and PCH-EM2 produced conversion gain uncertainties comparable to PT's in the $\sigma_R> 0.42,e\text{-}$ regime. This finding suggests that once the PCD's structure is lost due to increasing read noise, there is only a small advantage to using the full PCD model in the estimation process. Overall, PCH-EM performed the best for one-sample methods, while PCH-EM2 outperformed PCH-EM and showed potential as a general estimation technique in the sub-electron and multi-electron read noise regimes.

% Looking at only the one-sample methods, the PCH and Fourier methods beat PT in terms of RMSE below about $\sigma_R\approx 0.42\,e\text{-}$; however, above this read noise value they performed poorly because they both rely on peak detection. Unlike the PCH and Fourier methods, PCH-EM does not require observing peaks and is still able to estimate the conversion gain at larger read noise values; thus, bridging the gap into the multi-electron read noise regime for one-sample methods. In regards to the two-sample methods, both Nakamoto's method and PCH-EM2 produce conversion gain uncertainties on par with that of PT in the $\sigma_R> 0.42\,e\text{-}$ threshold.  This indicates that once the structure in the PCD is lost due to increasing read noise, there is only a small benefit to using the full PCD model in the estimation procedure. Overall, PCH-EM performed best for one-sample methods and PCH-EM2 performed best for the two sample methods. Furthermore, PCH-EM2 outperformed PCH-EM and shows promise as a general estimation procedure in the sub-electron and multi-electron read noise regimes.

%%%%%%%%%%%%%%%%%%%%%%%%%%%%%%%%%%%%%%%%%%%%%%%%%%%%%%%%%%%%%%%%
%%%%%%%%%%%%%%%%%%%%%%%%%%%%%%%%%%%%%%%%%%%%%%%%%%%%%%%%%%%%%%%%
%%%%%%%%%%%%%%%%%%%%%%%%%%%%%%%%%%%%%%%%%%%%%%%%%%%%%%%%%%%%%%%%

\section{Discussion and Future Work}
\label{sec:conclusions}

This study presented an overview of the PCD model as a universal framework for describing all currently available methods of conversion gain estimation in the sub-electron and multi-electron read noise regimes. By unifying the notation and model, it became possible to compare and contrast the differences between these methods. Monte Carlo experiments revealed that utilizing the full PCD model in the estimation procedure produced conversion gain estimates with less uncertainty than the traditional PT method, especially when the read noise is below $\sigma_R\approx 0.42\,e\text{-}$. Notably, the PCH-EM2 algorithm outperformed the time tested PT method below this threshold, while its performance merged with that of PT at higher read noise levels. This suggests that PCH-EM2 could potentially replace PT as a general estimation procedure. Future research will involve developing and implementing a multi-sample ($\geq 2$ sample) PCH-EM method and exploring the use of annealing to enhance the algorithm's robustness to poor starting points.

% In this work the PCD model was introduced and all of the available methods of conversion gain estimation in the sub-electron and multi-electron read noise regimes were unified under this model. The unified model and notation subsequently allowed for understanding the differences between each method. Monte Carlo experiments showed that when the read noise is below $\sigma_R\approx 0.42\,e\text{-}$ incorporating the full description of the PCD into the estimation procedure produced conversion gain estimates with less uncertainty than the traditional PT method.  In particular the PCH-EM2 algorithm outperformed PT below the $\sigma_R\approx 0.42\,e\text{-}$ threshold but merged with PT's performance at higher read noise suggesting it can potentially replace PT as a general estimation procedure. As such, future work will entail a full derivation of a multi-sample ($\geq 2$ sample) PCH-EM method. Additionally, there are opportunities to extend a multi-sample version of PCH-EM by incorporating annealing into the estimation procedure to make the algorithm more robust to poor starting points.

%%%%%%%%%%%%%%%%%%%%%%%%%%%%%%%%%%%%%%%%%%%%%%%%%%%%%%%%%%%%%%%%
%%%%%%%%%%%%%%%%%%%%%%%%%%%%%%%%%%%%%%%%%%%%%%%%%%%%%%%%%%%%%%%%
%%%%%%%%%%%%%%%%%%%%%%%%%%%%%%%%%%%%%%%%%%%%%%%%%%%%%%%%%%%%%%%%

% \appendix

% \section{First Appendix}
% \label{sec:xxx}

% Put appendix contents here if needed...

%%%%%%%%%%%%%%%%%%%%%%%%%%%%%%%%%%%%%%%%%%%%%%%%%%%%%%%%%%%%%%%%
%%%%%%%%%%%%%%%%%%%%%%%%%%%%%%%%%%%%%%%%%%%%%%%%%%%%%%%%%%%%%%%%
%%%%%%%%%%%%%%%%%%%%%%%%%%%%%%%%%%%%%%%%%%%%%%%%%%%%%%%%%%%%%%%%

\acknowledgments % equivalent to \section*{ACKNOWLEDGMENTS}       
 
The authors wish to express their gratitude to Nicholas Shade at Dartmouth University for his feedback on the implementation of the PCH method. The authors also would like to acknowledge and thank Katsuhiro Nakamoto from Hamamatsu Photonics for his help in implementing his method. Their contributions have been invaluable to the research, and the authors are appreciative of their assistance.

% References
\bibliography{report} % bibliography data in report.bib
\bibliographystyle{spiebib} % makes bibtex use spiebib.bst

\end{document}